\documentclass[10pt]{iopart}
\usepackage{iopams}  
\usepackage{mathrsfs} 
\usepackage{color}
\usepackage[usenames,dvipsnames,svgnames,table]{xcolor}
\usepackage{ulem}

\newcommand{\eqref}[1]{\eref{#1}}

\begin{document}

\title{Radiation Memory, Boosted Schwarzschild Spacetimes and Supertranslations}
\author{Thomas M\"adler$^1$\footnote{Email:tm513@cam.ac.uk} and Jeffrey Winicour$^{2,3}$}

\address{
${}^{1}$  Institute of Astronomy, University of Cambridge, Madingley Road, Cambridge, CB3 0HA, UK \\
${}^{2}$ Department of Physics and Astronomy \\
        University of Pittsburgh, Pittsburgh, PA 15260, USA\\
${}^{3}$ Max-Planck-Institut f\" ur
         Gravitationsphysik, Albert-Einstein-Institut, \\
	 14476 Golm, Germany \\
	}

\begin{abstract}
We investigate gravitational radiation memory and its
corresponding effect on the asymptotic symmetries of
a body whose exterior is a boosted Schwarzschild spacetime.
First, in the context of linearized theory, we consider such a Schwarzschild body which is initially at rest,
then goes through a radiative stage and finally emerges as a boosted Schwarzschild body. We show that
the proper retarded solution of the exterior Schwarzschild
spacetime for this process can be described in terms of the ingoing Kerr-Schild form
of the Schwarzschild metric
for both the initial and final states.
An outgoing Kerr-Schild or time symmetric metric does not give the proper solution.
The special property of Kerr-Schild metrics that their linearized and nonlinear forms are
identical allows us to extend this result to processes in the nonlinear regime. We then
discuss how the nonlinear memory effect, and its associated supertranslation,
affect angular momentum conservation. Our approach provides a new
framework for studying nonlinear aspects of the memory effect.
\end{abstract}


\maketitle

\section {Introduction}

Gravitational radiation memory is an effect which leads to the net displacement between test particles after the passage of a wave.
It was initially recognized in the treatment of dense stellar clusters in linearized gravitational theory,
where it arises from the burst of radiation accompanying the ejection of massive particles to infinity~\cite{zeld,brag}. 
This effect associated with the final boosted state of the ejected particles
was also found as a zero frequency radiation mode 
in studies of the collision of relativistic point particles~\cite{Smarr,BontzPrice}.
Later, a nonlinear form of radiation memory was  discovered by  Christodoulou~\cite{Christ_mem} 
and interpreted as a flow
of gravitons or gravitational wave energy to null infinity ~\cite{thorn,wisemanwill,frauen}.  
The relation between the linear and nonlinear versions of gravitational radiation
memory was clarified in a study of an analogous memory effect associated with electromagnetic
radiation in Maxwell theory~\cite{bieri1}.
Studies of the global sky pattern of radiation memory in both electromagnetic and
linearized gravitational theory have revealed its connection to the Maxwell gauge
freedom at null infinity and an analogous gravitational supertranslation 
freedom~\cite{globalemmem,skypattern}.
Further studies revealed that even in linearized gravitational theory
the energy flux of massless fields, e.g. electromagnetic fields, to null infinity generates gravitational radiation
memory~\cite{Bieri_null, Bieri_neutrino,tolish_wald,tolish}. 
For that reason, Christodoulou's nonlinear memory effect is commonly referred to as null memory.
Many other aspects of gravitational  memory have been discussed,
where those most relevant to our work
are~{\cite{hollands,emAsympt,mem_angular,extBMSCharge,HorizonMem,mem_soft_theorem,soft_info}.
Despite the stunning recent observation of gravitational  waves by the
LIGO-Virgo consortium~\cite{LIGO}, the detection of gravitational radiation memory is more demanding
due to the current insensitivity of the LIGO detectors and pulsar timing arrays~\cite{NANOGrav,pulsta}
to the long rise-time of typical memory signals~\cite{Favata}. 
In addition to the observational aspects of radiation memory,
there are compelling theoretical issues associated
with the induced supertranslation induced and its effect on angular momentum.
In this paper, we introduce a new method for exploring these issues.

The global sky pattern of gravitational radiation memory
can be decomposed into $E$  and $B$  modes~\cite{globalemmem,skypattern},
analogous to the classification of electromagnetic radiation, {e.g. of the
pattern of the cosmic microwave background radiation~\cite{bmode}.}
In the electromagnetic case \cite{globalemmem}, the radiation field
at retarded time $u$ at future null infinity
$\mathcal{I}^+$ can be described in standard angular coordinates  $x^B = (\theta,\phi)$ by the electric field
$E_A(u,x^B)$. (The angular components of the radiation field are finite at $\mathcal{I}^+$.)
The memory effect produces a net momentum kick $\Delta  p_A$ on a test charge $q$ after the radiation passes,
\begin{equation}
\Delta  p_A = q \int_{u=-\infty}^{u=+\infty}  E_A(u,x^B) du .
 \label{eq:emem}
 \end{equation}
Here we introduce the notation
\begin{equation}
\Delta F(x^C):=F(u,x^C)|_{u=\infty}-F(u,x^C)|_{u=-\infty}.
\end{equation}
The $E$ and $B$ mode components, $E_{A[e]}$ and $E_{A[b]}$, respectively,
result from decomposing the electric field in terms of a gradient and the dual of a gradient,
\begin{equation}\label{dual_decomp}
           E_A =E_{A[e]}+ E_{A[b]} = \eth_A \Phi_{[e]} + \epsilon_{CA} \eth^C \Phi_{[b]},
\end{equation}
where $\eth_A$ and $\epsilon_{AB}$ are the covariant derivative and surface area
tensors with respect to the unit sphere
metric $q_{AB}$.  {This is the 2-dimensional version of the Helmholtz decomposition
of a vector field into a gradient and curl.} The scalar
fields  $\Phi_{[e]}$ and $\Phi_{[b]}$ determine the E-mode and B-mode, respectively.
This decomposition is distinct from the separation of the Maxwell field into
electric and magnetic radiation fields, which are related by $|\vec E| =|\vec B|$
and $\vec E \perp \vec B$. The electromagnetic radiation pattern and its memory could be equally
well described in terms of its $\vec B$ field.

A compact way to describe the $E$ and $B$ mode decomposition is in terms of
a complex polarization vector $q_A$ satisfying 
\begin{equation}
          q _{AB} = q_{(A} \bar q_{B)}, \quad \epsilon_{AB} =i q_{[A} \bar q_{B]},
            \quad q^A \bar q_A =2, \quad q^A q_A =0,
\end{equation}
e.g. $q_A = (1, i \sin \theta)$ for the standard form of the unit sphere metric $q_{AB}$.
(Here $(AB)$ and $[AB]$ denote, respectively, symmetrization and antisymmetrization
of indices.)
Then the electromagnetic radiation is represented by the complex spin-weight-1 field
\begin{equation}\label{complexModeDecomp}
          q^A E_A = q^A (\eth_A \Phi_{[e]} +\epsilon_{CA} \eth^C \Phi_{[b]}) = \eth(\Phi_{[e]}+i \Phi_{[b]}),
\end{equation}
where $\eth$ is the Newman-Penrose spin-weight raising operator~\cite{BMS2,goldberg_eth,penrin,np_scholar}.

{The memory due to charged particles escaping to infinity (or their time reversed capture) and the null
memory due to the flow of  a hypothetical massless charged field or
massless charged particles to future null infinity ${\mathcal I}^+$ were discussed in~\cite{bieri1}.}
There is another mechanism for electromagnetic radiation memory due
to source-free ingoing-outgoing waves ~\cite{globalemmem}.

In \cite{globalemmem} it was shown, in the absence of magnetic monopoles and with other
reasonable physical assumptions, that only source-free
waves produce B-mode memory. This result was obtained by an analysis
of Maxwell's equations in a null gauge, analogous to the Bondi-Sachs
formalism for the gravitational field. In terms of
outgoing null coordinates $x^\alpha=(u,r,x^A)$, $u=t-r$, the Maxwell field $F_{\alpha\beta}$ is
expressed in terms of a vector potential $A_\alpha$ as
$F_{\alpha\beta} =\partial_\alpha A_\beta -\partial_\beta A_\alpha$, whose
gauge freedom is used to set 
$ A_r(u,r,x^C)=0$ and $A_u(u,\infty,x^C) =0$.
The remaining gauge freedom, 
\begin{equation}
   A_B \rightarrow A_B +\eth_B\Lambda(x^C) ,
\label{eq:egauge}
\end{equation}
is purely $E$-mode. It is the analogue of the supertranslation freedom
in the Bondi-Metzner-Sachs (BMS) asymptotic symmetry group \cite{bondi,sachs,Sachs_BMS,tam}
in the gravitational case. For tutorials on the Bondi-Sachs formalism
see the Scholarpedia article~\cite{bs_scolar} or the Living Review article~\cite{winrev}.

In this gauge, the radiation field at ${\mathcal I}^+$ is given by $E_B(u,x^A) = -\partial_u A_B(u,x^A)$,
where $A_B$ has an $E$ and $B$ mode decomposition similar to \eqref{dual_decomp}.
The radiation memory $\Delta p_A$ for a unit charge (\ref{eq:emem}) then reduces to
\begin{equation}
\int_{-\infty}^{\infty}   E_B(u,x^A) du |_{{\mathcal I}^+} =
       - \Delta A_B(x^A)   |_{{\mathcal I}^+}.
\end{equation}}
Except in the case of $B$-mode homogeneous wave memory,
this is an effective $E$-mode gauge shift between $u=\pm \infty$, i.e.
the remaining gauge freedom (\ref{eq:egauge}) allows either $A_B(\infty, x^A) |_{{\mathcal I}^+}$
or $A_B(-\infty, x^A) |_{{\mathcal I}^+}$ to be gauged to zero but not their difference.
As an example, the boost memory from a charge $Q$ initially at rest
and then ejected with velocity $V$ in $z$-direction
is given by 
\begin{equation}
    \Delta  A_B|_{{\mathcal I}^+}= \frac {Q V}{1-V \cos\theta} \eth_B \cos\theta ,
\end{equation}
which corresponds to a gauge shift with
\begin{equation}
\label{em_gauge_shift}
\Lambda= Q \ln(1-V\cos\theta).
\end{equation}

In \cite{skypattern}, the analogues of these results for electromagnetic radiation memory
were established for linearized gravitational theory, where the radiation is described by a trace-free strain tensor
at ${\mathcal I}^+$,
\begin{equation}
      \sigma_{AB}(u,x^C) , \quad q^{AB}  \sigma_{AB} =0.
\end{equation}
{In the notation of~\cite{thorn,wisemanwill}, $ \sigma_{AB}$ corresponds to the description of the
gravitational wave by the asymptotic limit of $r h^{TT}_{ij}$, determined by
the trace-free component of the metric transverse to the observer's
line of sight.   }

The radiation memory is determined by the change in strain $\Delta\sigma_{AB}(x^C)$.
The $E$ and $B$ mode decomposition of the gravitational radiation field can be made by first noting
that the strain can be represented in terms of a deformation vector field $\xi^A$  by
\begin{equation}\label{def_strain_sigma_AB}
    \sigma_{AB}=\eth_{(A} \xi_{B)} -\frac {1}{2} q_{AB} \eth^C \xi_C .
\end{equation}
Given two real scalar fields $\Sigma_{[e]}$ and $\Sigma_{[b]}$ that form the complex
scalar field $\Sigma:=\Sigma_{[e]}+i\Sigma_{[b]}$, this can be further decomposed as
\begin{equation}
     \xi_A = \eth_A \Sigma_{[e]} + \epsilon_{BA} \eth^B \Sigma_{[b]}.
\end{equation} 
In terms of the Newman-Penrose spin-weight formalism,
this leads to the spin-weight-2 representation of the strain,
\begin{equation}
     \sigma:=q^A q^B \sigma_{AB}=q^A q^B \eth_A \eth_B\Sigma =\eth^2 \Sigma \, ,
       \quad \Sigma = \Sigma_{[e]}+i\Sigma_{[b]} .
\end{equation}
Here $\Sigma$ is the spin-weight-0
potential generating the spin-weight-2 field $\sigma$ via the  spin-weight raising operator $\eth$.
In spin-weight terminology, $\Sigma_{[e]}$ and $\Sigma_{[b]}$ represent the 
``electric'' and ``magnetic'' parts of the strain, corresponding to the E and B
radiation modes. 
Similar to the electromagnetic fields $\vec E$ and $\vec B$, the gravitational
fields ${\bf E}$ and ${\bf B}$ 
picked out from the Weyl curvature tensor by a timelike vector $T^a$
are equal in magnitude and rotated by $45^o$.
The $E$ and $B$ mode decomposition is a separate global distinction.

Analogous to the electromagnetic case,
there are three known mechanisms for gravitational radiation memory:
\begin{itemize}

\item Boost memory due to the ejection of massive particles to infinity.
{In~\cite{thorn},
the boost memory is interpreted in terms of the trace-free-transverse component
of the $1/r$ Coulomb-type field corresponding to the 4-momentum of the particles.}
     
\item  Null memory due to the loss of energy to ${\mathcal I}^+$ 
by electromagnetic radiation, or radiation loss by other massless fields including gravitational
radiation in the nonlinear regime (Christodoulou memory).

\item Homogeneous wave memory due to an ingoing-outgoing gravitational wave propagating
from ${\mathcal I}^-$ to ${\mathcal I}^+$.

\end{itemize}

In \cite{skypattern} it was shown, under physically reasonable asymptotic conditions
on the matter fields, that only homogeneous gravitational
waves produce B-mode memory. Conventional matter sources can produce
$B$-mode gravitational waves, but they have vanishing radiation memory.
Thus $B$-mode radiation memory must be of essentially primordial origin, corresponding
in linearized theory to a source-free wave entering from ${\mathcal I}^-$.

As discussed in~\cite {globalemmem,skypattern,mem_soft_theorem},
gravitational radiation memory is related to the supertranslation freedom
$u\rightarrow u+\alpha(x^C)$ in the
choice of retarded time for an asymptotic inertial frame at ${\mathcal I}^+$. During a non-radiative
period, in which $\partial_u \sigma =0$, the strain has the
supertranslation gauge freedom
\begin{equation}
\label{supertranslation}
\sigma (x^C)\rightarrow \sigma(x^C)+\eth^2 \alpha(x^C).
\end{equation} 
Thus the E-mode
component of the strain can be gauged away during a non-radiative epoch.
Because the supertranslation $\alpha$ is a real function it does not affect the B-mode.
The E-mode radiation memory $\Delta\sigma_{[e]}$ is gauge invariant and can be considered to
be a supertranslation shift between the two preferred gauges that the strain
picks out at $u=\pm \infty$. For a body which is initially
at rest and then ejected with mass $m$ and velocity $V$ in the $z$-direction, the radiation memory is
(see Eq.~\eqref{eq:bstrain})
\begin{equation}
\Delta\sigma  =\frac{4m\Gamma V^2\sin^2\theta}{1-V\cos\theta},
\label{eq:mem1}
\end{equation}
{which agrees with Eq. (1) of~\cite{thorn}.}
{The supertranslation corresponding to (\ref{eq:mem1}) is}
\begin{equation}\label{eq:sup1}
\alpha=4m \Gamma(1-V\cos\theta)\ln(1-V\cos\theta) .
\end{equation}
{Note that it is the pure E-mode form of the radiation memory (\ref{eq:mem1})
that allows it to be interpreted in terms of a single real function $\alpha$ as a
supertranslation shift.}
It is curious, and perhaps of some deeper significance, that this has an
electromagnetic 
analogue in the E-mode gauge shift \eqref{em_gauge_shift} of the
vector potential.

The supertranslation freedom in the BMS
group has bearing on the treatment of the angular momentum
of an isolated gravitational system~\cite{BMS2,gw, jwangular}. The BMS group
consists of the semi-direct product of the infinite dimensional supertranslation
group with the Lorentz group. The 4-parameter time and space translation subgroup,
picked out by supertranslations $\alpha(\theta,\phi)$ composed of spherical harmonics
with $\ell\le 1$, form an
invariant subgroup, which leads to an unambiguous
definition of energy-momentum.  Although the Lorentz group is also
a subgroup of the BMS group, the supertranslations lead to a mixing of
the associated supermomentum with angular momentum. We discuss in
Sec.~\ref{sec:ang} how this supertranslation freedom can be fixed during
a non-radiative epoch and the BMS group reduced to the Poincar{\' e} group, in which
case angular momentum can be uniquely defined.
However, for a system
which makes a non-radiative to non-radiative transition, the two Poincar{\'e}
groups obtained at early and late times are shifted by a
supertranslation. Such supertranslation shifts associated with
radiation memory complicate the
treatment of angular momentum.  

In this paper, we develop a model based upon a Kerr-Schild version of the
Schwarzschild metric for studying
the production of radiation memory and its effect on the BMS group.
We use the Kerr-Schild metric to describe the static, spherically symmetric
spacetime  exterior to what we refer to as a Schwarzschild body. 
First, in Sec.~\ref{sec:lin},  we consider such a Schwarzschild body
in linearized theory which is initially at rest,
then goes through a radiative stage and finally emerges as a boosted Schwarzschild body. We show that
the proper retarded solution of the exterior
spacetime for this process is described in terms of the ingoing version of
the Kerr-Schild metric for both the initial and final states.
An outgoing Kerr-Schild or time symmetric Schwarzschild metric does not give the proper solution.
Then, in Sec.~\ref{sec:nlin}, we extend this model to processes in the nonlinear regime
by using the special property of the Kerr-Schild metrics that their linearized and nonlinear versions are
identical. We discuss the resulting nonlinear memory effect, its associated supertranslation and  the details
of its effect on the treatment of angular momentum.

Although an exact Schwarzschild exterior is
unrealistic in a dynamic spacetime it is reasonable to expect that our results
hold if it is a good approximation in the neighborhood of spatial infinity (infinite
past retarded time) and timelike infinity (infinite future retarded time). The underlying
justification is that a finite radiative energy loss to ${\mathcal I}^+$ requires
that $\partial_u \sigma(u,x^A)\rightarrow 0$ as $u\rightarrow \pm \infty$.  Furthermore,
the Kerr-Schild model has a natural extension to the case where the final state in the
neighborhood of timelike infinity is approximated by a Kerr black hole. 


\section{Linearized memory of boosted Schwarzschild transitions}
\label{sec:lin}
In linearized theory consider a compact spherically symmetric
ball of matter. Its exterior spacetime is described by a linearized  Schwarzschild
perturbation $h_{ab}$ of the Minkowski metric $\eta_{ab} =diag(-,+,+,+)$ in Cartesian inertial
coordinates $x^a= (t,x^i) =(t,x,y,z)$.
We consider three versions of this exterior Schwarzschild perturbation $h_{ab}$
in coordinates adapted to the static Killing vector $\partial_t$.
 
 \begin{itemize}
 
 \item The isotropic time symmetric version 
 \begin{equation}
h_{ab}^{(i)} = \frac{2M}{r}\eta_{ab}+\frac{4M}{r}T_aT_b ,
\end{equation}
where $T_{a} = -\partial_a t$.

 \item The outgoing Kerr-Schild form 
\begin{equation}
h_{ab}^{(k)} = \frac{2M}{r}k_a k_b,
\end{equation}
where $k_a=T_a + r_a$ and $r_a=\partial_a r$, with $r^2 = \delta_{ij}x^ix^j$.
Here $k_a$ is the outgoing radial null vector satisfying $k_a =-\partial_a u$ in terms of the retarded time $u=t-r$.
 
 \item The ingoing Kerr-Schild form
 \begin{equation}
h_{ab}^{(n)} = \frac{2M}{r}n_a n_b.
\end{equation}
 where $n_a=T_a-r_a$ is the
ingoing radial null vector satisfying $n_a =-\partial_a v$ in terms of the advanced time $v=t+r$.
 \end{itemize}
 These three perturbations differ from each other by a gauge transformation. 
They all have zero linearized radiation strain at both ${\mathcal I}^+$ and ${\mathcal I}^-$.

Now consider the boosted versions of these perturbations $h_{ab}^{(bi)}$, $h_{ab}^{(bk)}$ and $h_{ab}^{(bn)}$,
corresponding to a 
Schwarzschild body with 4-velocity $v^a = \Gamma(1, V^i)$, where $\Gamma = (1 -\delta_{ij}V^iV^j)^{-1/2}$.
With respect to the initial
rest frame, the boosted version can be obtained by
the substitutions $T^a \rightarrow v^a$, $k_a \rightarrow K_a = v_a+R_a$,  $n_a \rightarrow N_a = v_a-R_a$,
\numparts
\begin{equation}
r^2\rightarrow R^2=  x_a x^a + (x_av^a)^2
     = r^2\Big[1-\frac{t^2}{r^2}+ \Gamma^2\Big(r_iV^i - \frac{t}{r}\Big)^2\Big]
\end{equation}
and 
\begin{equation}
r_a \rightarrow  
R_a =\frac{x_a  + (x_bv^b)v_a}{R}
     = \frac{x_a  - r\Gamma(\frac{t}{r}-V^ir_i)v_a}{R} \, ,
\end{equation}
\endnumparts
which reduce to the rest-frame  expressions if $V^i=0$.

Again, these boosted perturbations all differ by a gauge transformation.
They are all non-radiative so that they have zero radiation
memory. But, relative to the unboosted rest frame, they have (pure gauge) radiation strain: 
either non-zero asymptotic strain $\sigma_{+}$ at ${\mathcal I}^+$,
or non-zero asymptotic strain $\sigma_{-}$ at ${\mathcal I}^-$, or both.

In order to construct the boosted radiation strain relative to the initial rest frame we introduce a
complex polarization vector $Q^a$ satisfying
\begin{equation}
\eta_{ab}Q^a\bar Q^b -2= \eta_{ab}Q^a Q^b = 0\;\;,\;\;\;
Q^aT_a=Q^a r_a =Q^a x_a= 0.
\end{equation}
Here $Q^a$ has the components
\begin{equation}
Q^a = (0, \cos\theta\cos\phi-i\sin\phi,
             \cos\theta\sin\phi+i\cos\phi,-\sin\theta)
\end{equation}
and is the Cartesian analogue of the polarization dyad $q^A=(1, i\sin^{-1}\theta)$ on
the unit sphere, chosen so that its angular components satisfy $Q^A=r^{-1}q^A$.
The linearized radiation strain is given by
\begin{equation}\label{eq:def_strain+-}
\sigma_{\pm}=\frac{r}{2} Q^aQ^b h_{ab}|_{{\mathcal I}^{\pm}} .
\end{equation}

In order to calculate $\sigma_+$ at ${\mathcal I}^+$, for which 
$r\rightarrow \infty$ holding $u$ constant, we express $R^2$ and $R_a$  as functions of $(u,r,x^A)$,
\numparts
\begin{eqnarray}
R^2  &=& r^2\Big[-\frac{u^2}{r^2}-2\frac{u}{r}+ \Gamma^2\Big(1-r_iV^i + \frac{u}{r}\Big)^2\Big]\;\;, \\
R_a &=& \frac{x_a  - r\Gamma(\frac{u}{r}+1-V^ir_i)v_a}{R} \;\;.
\end{eqnarray}
\endnumparts
Similarly, for the calculation of $\sigma_-$, we express $R^2$ and $R_a$  as functions of $(v,r,x^A)$,
\numparts
\begin{eqnarray}
R^2  
  &=& r^2\Big[-\frac{v^2}{r^2}+2\frac{v}{r}+ \Gamma^2\Big(1+r_iV^i - \frac{v}{r}\Big)^2\Big]\;\;,\\
R_a      & = &\frac{x_a  + r\Gamma(-\frac{v}{r}+1+V^ir_i)v_a}{R}\;\;.
\end{eqnarray}
\endnumparts
As functions of $(u,r,x^A)$, the scalar products $Q^aK_a =Q^a(v_a+R_a) $  and $Q^aN_a = Q^a(v_a-R_a)$
{have limits at ${\mathcal I}^+$}
{
\numparts
 \begin{eqnarray}
\lim_{\stackrel{r\rightarrow\infty}{u=const}} Q^a K_a
&=&  0\;\;, \\
\lim_{\stackrel{r\rightarrow\infty}{u=const}} Q^a N_a  
&=&  2\Gamma Q^iV_i\;\;,
\end{eqnarray}
\endnumparts
}
and as function of $(v,r,x^A)$ {have limits at ${\mathcal I}^-$}
{
\numparts
\begin{eqnarray}
\lim_{\stackrel{r\rightarrow\infty}{v=const}} Q^a K_a 
&=&  2 \Gamma Q^iV_i \;\;,\\
\lim_{\stackrel{r\rightarrow\infty}{v=const}} Q^a N_a   
&=& 0 \, .
\end{eqnarray}
\endnumparts
}

Using these limits, we obtain the radiation strains for the three boosted perturbations,
\numparts
\begin{equation} 
     {\sigma_+}^{(bi)} = \frac{2M\Gamma( Q^iV_i)^2}{1-r_iV^i} \, , \quad 
      {\sigma_-}^{(bi)} = \frac{2M\Gamma( Q^iV_i)^2}{1+r_iV^i} \;\;,
\end{equation}
\begin{equation} 
     {\sigma_+}^{(bk)} = 0\, , \quad \quad\quad\quad\quad\quad\,
      {\sigma_-}^{(bk)} = \frac{4M\Gamma( Q^iV_i)^2}{1+r_iV^i} \;\;,
\end{equation}
\begin{equation} 
     {\sigma_+}^{(bn)} = \frac{4M\Gamma( Q^iV_i)^2}{1-r_iV^i} \, , \quad 
      {\sigma_-}^{(bn)} =0 \, . 
\end{equation}
\endnumparts
Now we return to the perturbation calculation of the radiation memory
at ${\mathcal I}^+$ produced by a Schwarzschild body which
is initially at rest, then goes through a radiative period described by a retarded
solution and ends up as a boosted Schwarzschild body.
Relative to the initial rest frame, the initial radiation strain vanishes 
for all three perturbations, i.e. $\sigma_+(u=-\infty, x^A)=0$.
For the three perturbations, we then have
\numparts
\begin{eqnarray}
    \Delta {\sigma_+}^{(bi)} &=& \frac{2M\Gamma( Q^iV_i)^2}{1-r_iV^i} \, , \\
     \Delta {\sigma_+}^{(bk)} &=&0\;, \\
     \Delta {\sigma_+}^{(bn)} &=& \frac{4M\Gamma( Q^iV_i)^2}{1-r_iV^i}\;\; .         \label{eq:sigma} 
\end{eqnarray}
\endnumparts
In comparison with the memory computed by the retarded  solution given in~\cite{brag},
only $\Delta {\sigma_+}^{(bn)}$, the memory associated with the ingoing Kerr-Schild
perturbation, gives the correct result. (Note that equation (2b) of~\cite{brag} for the $\times$
component of memory is correct but the right hand side of equation (2a) for the $+$ component
should be increased by a factor of 2. {Furthermore, for a particle ejected (captured)
in the $z-$direction, \eqref{eq:sigma} agrees with equation (1) of \cite{thorn}  when using \eqref{eq:def_strain+-}
to define the line-of-sight components  $h^{TT}_{ij}$ .)}

Energy-momentum would not be conserved in an initial state consisting of a body
of mass $M$ at rest and a final state consisting of a boosted body with mass $M$.
However, given the same initial state it is conserved by
a final state consisting of two bodies of mass $m$ moving with
velocities $\pm V^i$, where $M=2\Gamma m$. (Here we note that there
is no energy loss due to gravitational radiation in the linearized approximation.)
In that case, by superposition, the resulting radiation memory for the ingoing
Kerr-Schild model is 
\begin{equation} 
           \Delta {\sigma_+}^{(bn)} = \frac{4M( Q^iV_i)^2}{1-(r_iV^i)^2}.
           \label{eq:memsplit}
\end{equation}


\section{Nonlinear memory}
\label{sec:nlin}

The linearized treatment of radiation memory in Sec.~\ref{sec:lin} can be generalized to many different
scenarios by means of the superposition principle, e.g.
to the capture or scattering of a particle or system of particles. Since such linearized
processes could also be treated by standard methods, we turn our
attention here to the nonlinear case. In doing so we take advantage of two special properties
of the Kerr-Schild form of the Schwarzschild solution: (i) the nonlinear solution is identical to the solution
linearized with respect to $m$ and (ii) the Minkowski background intrinsic to
the Kerr-Schild geometry
provides a natural way to introduce Poincar{\' e} transformations and, in particular,
boosts. Of course, the nonlinear case does not allow superposition of
Kerr-Schild solutions except in the approximation where
the masses are very far apart. (See~\cite{mkerrsch} for a discussion
of posing multiple black {hole} data by means of superimposed Kerr-Schild black holes.)

The results for the linearized case show that the proper treatment of the radiation
memory corresponding to a retarded solution must be based upon a Schwarzschild body
whose exterior is described by the ingoing version
of the Kerr-Schild metric, which in the rest frame is given by
\begin{equation}
   g_{ab} = \eta_{ab} + \frac{2M}{r} n_a n_b .
\end{equation} 
This exterior metric has vanishing radiation strain.  We consider two nonlinear scenarios.

\subsection{Scenario 1}
\label{sec:s1}

Consider first the nonlinear version of the linearized system in Sec.~\ref{sec:lin}
consisting initially of a Schwarzschild body of mass $M$ at rest which, after some radiative interval,
ends up in the final state of two boosted Schwarzschild bodies of mass $m$ separating with velocities
$\pm V^i$. The linearized result should be expected to extend reliably
to the nonlinear regime in the {approximation
where the energy loss due to gravitational radiation is small}, so that
$M=2m\Gamma$. 
{For boosts in the $\pm z-$direction with $V_z =\pm V$, the resulting
radiation memory at ${\mathcal I}^+$, as given by (\ref{eq:memsplit}), reduces to}

\begin{equation}
          \Delta \sigma = \frac{4MV^2 \sin^2 \theta}{1- V^2 \cos^2\theta}.
\end{equation} 
{In the axisymmetric case, where
$$q^A q^B \eth_A \eth_B \alpha=\eth^2 \alpha 
=\sin\theta \partial_\theta (\sin^{-1}\theta \partial_\theta \alpha),
$$
the corresponding supertranslation shift 
obtained from (\ref{supertranslation}) is
\begin{equation}
     \alpha = -2M \Big[(1-V\cos\theta) \ln (1-V\cos\theta)+(1+V\cos\theta) \ln (1+V\cos\theta) \Big] .
\end{equation}
}

In the foregoing approximation, there is
negligible energy loss due to gravitational radiation.
But this model can be extended to the case
of a radiative mass loss $\Delta E = M-2m\Gamma$ during the period of acceleration
leading to the boost.
In that case, the contribution to the radiation memory due to the boost is
\begin{equation}
          \Delta \sigma = \frac{8m\Gamma V^2 \sin^2 \theta}{1- V^2 \cos^2\theta}.
\end{equation}
Note, however, that the null (Christodoulou) memory associated with the radiative
loss $\Delta E$ is independent of the boost memory. We discuss this in more detail
in the next scenario.

\subsection{Scenario 2}
\label{sec:s2}

In the nonlinear context, it is consistent with energy-momentum conservation to consider
the  simple scenario of an initial Schwarzschild body of mass $M$ at rest which, after a radiative interval, leads a final state
of mass $m$ with boost velocity $V$ in the $z$-direction. From (\ref{eq:sigma}),
the resultant boost memory $\Delta \sigma_{[B]}$ is
\begin{equation}
    \Delta \sigma_{[B]} = \frac{4m \Gamma V^2 \sin^2 \theta}{1- V \cos\theta} ,
     \label{eq:bstrain}
\end{equation}
with the associated supertranslation (\ref{eq:sup1}).
The radiative energy-momentum loss
is given by the future pointing timelike vector,
\begin{equation}
        \Delta P^\alpha = (\Delta E, \Delta P^i) 
        = \frac{1}{4\pi}\int_{-\infty}^{\infty} du \oint  \|N\|^2 \ell^\alpha \sin\theta d\theta d\phi, 
\label{eq:ploss}
\end{equation}
where  $N=\partial_u \sigma$ is the Bondi news function and 
\begin{equation}
\ell^\alpha = (1,  \sin\theta \cos\phi, \sin\theta\sin\phi, \cos\theta) \;\;.
\end{equation}

{The associated null (Christodoulou) memory $\Delta \sigma_{[N]}$  depends
non-locally on the radiated energy flux.
The decomposition of the memory into a boost and null 
component is best described in the formulation of Frauendiener~\cite{frauen},
who finds in the vacuum case
\begin{equation}
    \eth^2  \Delta \bar \sigma = -\Delta\psi_2^{[0]} 
            +\int_{-\infty}^{\infty} \| N \|^2 du,
            \label{eq:gmem}
\end{equation}
where $\psi_2^{(0)}$ is the leading asymptotic part of the
Newman-Penrose Weyl tensor component~\cite{NP}.
The first term in (\ref{eq:gmem}) is, in the intuitive picture
of~\cite{thorn}, the
leading ``Coulomb'' part of the curvature and incorporates the boost
memory produced by freely escaping massive particles
in linearized theory.
The second term vanishes in linearized theory and represents
the null memory due to the gravitational
radiation of energy to ${\mathcal I}^+$,
as originally found by Christodoulou.
}

Energy-momentum conservation requires
\begin{equation}
     M= m\Gamma  + \Delta E,  \quad m\Gamma V= \Delta P^z .
\end{equation}
In the non-radiative regime where $\Delta E$ is small it follows that  $\Delta P^z$ and consequently
$V$ must also be small. In the other extreme, where $V$
is large, the timelike property of $\Delta P^\alpha$ requires
$\Delta E \ge \Delta P^z=m\Gamma V$ so that
$M \ge m\Gamma (1+V)$. As a result,
$$m\le M\sqrt{ \frac{1-V}{1+V}}$$
and
 \begin{equation}
    \Delta \sigma_{[B]} = \frac{4m \Gamma V^2 \sin^2 \theta}{1- V \cos\theta} 
    \le \frac{4M V^2 \sin^2 \theta}{(1+V)(1- V \cos\theta)} .
\end{equation}
In this way, energy-momentum conservation constrains
the boost component of
radiation  memory.

In other respects, $\sigma_{[B]}$ and $\sigma_{[N]}$ are independent.
A radiation burst of compact support in retarded time $u$ produces
null memory independent of whether there is any boost memory (ejected particles).
Conversely, boost memory can be produced with with negligible
radiative energy flux, i.e. negligible null memory.
{As an example, consider a particle which is at rest in the infinite
past and then boosted in the $z$-direction with final velocity $V$,
according to the retarded time dependence 
$$ z(u) = \frac{V}{\lambda} \ln \bigg(\frac{1+e^{\lambda u}}{2}\bigg),
$$
with velocity
$$ \dot z(u) = \frac{V}{1+e^{-\lambda u}}
$$
and acceleration
$$ \ddot z(u) = \frac{ \lambda Ve^{-\lambda u}}{(1+e^{-\lambda u})^2},
$$
where $\lambda >0$ controls the acceleration rate.
The quadrupole moment $Q\sim z^2$ has second derivative
$\ddot Q \sim z\ddot z + \dot z^2$, which has asymptotic 
behavior $\ddot Q|_{u=-\infty}=0$ and $\ddot Q|_{u=\infty}\sim V^2$.
As a result, in the quadrupole approximation, the corresponding radiation strain
leads to nonzero boost memory $\Delta \sigma_{[B]} \sim \Delta \ddot Q \sim V^2$.
However, for small $\lambda$ the news function has dependence
$N \sim\;  \stackrel{\boldsymbol{\ldots}}{Q}\; =\;O(\lambda)$ so that as $\lambda\rightarrow 0$
there is negligible radiation flux and negligible null memory.
This example underlies the intuitive picture that boost memory corresponds to the infrared
limit of the radiation.
 }

Note that in both scenario 1 and 2, the identification of radiation memory with a supertranslation
shift only depends on a transition between initial and final states with vanishing E-mode radiation.
This appears to be related,
in a way that is not completely understood,
to the result that B-mode radiation memory is only possible
for a source-free ingoing-outgoing wave.

\subsection{Angular momentum}
\label{sec:ang}

Although the example consider in Sec.~\ref{sec:s2} is not physically realistic,
it provides a simple framework for discussing
the effect of supertranslations on angular momentum loss.
Referring to (\ref{supertranslation}), the supertranslation
associated with (\ref{eq:bstrain}) is
\begin{equation}
    \alpha=-4m \Gamma(1-V\cos\theta)[\ln(1-V\cos\theta) -1] .
    \label{eq:alpha}
\end{equation}    

The initial stationary state picks out inertial coordinates $(u,x^A)$ on ${\mathcal I}^+$ (also called Bondi coordinates).
These initial coordinates extend uniquely to an inertial frame on all of ${\mathcal I}^+$.
In these coordinates, the 
generators of the BMS group have the form
\begin{equation}
  \xi^a \partial_a =\bigg [ \frac{u}{2}\eth_A f^A +{\mathcal A}(x^B)\bigg ]\partial_u +f^A(x^B)\partial_A \, ,
\end{equation}
where $\eth^{(A}f^{B)}=\frac{1}{2}q^{AB}\eth_C f^C$, i.e. $f^A$ is a conformal Killing vector
on the unit sphere. The supertranslation subgroup corresponds to $f^A=0$, where ${\mathcal A}$
is composed of $\ell\le 1$ spherical harmonics for the
time and space translations.
The subgroup ${\mathcal A}=0$ corresponds to a Lorentz subgroup with generators
\begin{equation}
  \eta^a \partial_a =\frac{u}{2}\eth_A f^A \partial_u +f^A \partial_A .
  \label{eq:eta}
\end{equation}
A fixed choice of ${\mathcal A}$ determines the supertranslation freedom in the choice of Lorentz subgroup.
The Lorentz transformations with $\eth_A f^A =0$ determine a rotation subgroup.
As described in~\cite{sflux}, we can set $f^A=\eth^A \Psi +\epsilon^{AB}\partial_B \Phi$, where $\Psi$ represents
the boost component and $\Phi$ is the rotation component. Boosts $B^{[i]a}$ corresponding to the Cartesian axes
are then determined by $\Psi^{[i]} =\ell^i$, in terms of the direction cosines  (\ref{eq:ploss});
and $\Phi^{[i]} =\ell^i$ for the corresponding rotations $R^{[i]a}$. Because $\ell^i$
consist of $\ell=1$ spherical harmonics, $\eth_A B^{[i]A} =\eth_A \eth^A \ell^i =-2 \ell^i$ and (\ref{eq:eta}) reduces to
\begin{equation}
  B^{[i]a} \partial_a = -u \ell^i \partial_u +(\eth^A \ell^i) \partial_A 
  \label{eq:bi}
\end{equation}
and
\begin{equation}
   R^{[i]a} \partial_a = \epsilon^{AB}(\eth_B \ell^i) \partial_A .
  \label{eq:ri}
\end{equation}
The 4-parameter translation subgroup has one time translational generator $A$ and three spatial
translational generators $A_i$ such that
\begin{equation}
 \tau^a \partial_a = (A+A_i \ell^i)\partial_u.
 \label{eq:ti}
\end{equation}

In the scenario depicted in Sec.~\ref{sec:s2}, there are various options for a preferred rotation subgroup.
In this simple model, one could pick out the initial (final) rotation subgroup to be the rotational symmetry
group of the initial (final) Schwarzschild solution. But this option would not be possible in a more general
setting where the final state consisted of more than one Schwarzschild bodies with different
boosts. Instead, we consider choices which only depend upon the $O(1/r)$  radiation
field, i.e. the asymptotic strain $\sigma$. This leads to the following two choices.

\begin{enumerate}

\item The initial state picks out inertial coordinates $x^a=(u, x^A)$,
where the surfaces $u =constant$ are strain-free. The BMS
transformations which leave the strain zero determine a preferred Poincar{\' e} 
subgroup. For a given choice of $u$, this leads to a preferred rotation
subgroup with generators satisfying $R^a\partial_a u=0$, i.e. with generators
tangent to the surfaces $u=const$.
For this choice of strain-free inertial coordinates, the Poincar{\' e} generators $B^a$,
$R^a$ and $\tau^a$ are given by (\ref{eq:bi}),  (\ref{eq:ri}) and (\ref{eq:ti}).
This initial choice of coordinates  $x^a=(u, x^A)$ and Poincar{\' e} generators
extend uniquely to all of ${\mathcal I}^+$.

In this construction, there remains translation and boost freedom in the initial choice of
inertial coordinates $(u,x^A)$. The strain-free condition reduces 
the supertranslation freedom $u\rightarrow u+\alpha$ to a translation,
which fixes the ``origin'' of the rotation group up to the freedom in special relativity.
The boost freedom could be fixed by tying the choice of $u$
to the initial 4-momentum $P^\alpha$. However, this option does not play a
role in the present discussion.
 
\item The final state picks out inertial coordinates $\tilde x^a=(\tilde u=u+\alpha, \tilde x^A =x^A)$,
where the surfaces $\tilde u =constant$ have vanishing strain. Here
the angular coordinates and their associated Cartesian axes designated by the direction
cosines $\ell^i$ have been chosen to be those of the initial inertial frame. The final state picks out
a preferred Poincar{\' e} group consisting of BMS transformations which maintain the zero strain
of the $\tilde u =constant$ surfaces.
The supertranslation group is thus restricted to the 4-parameter translation subgroup with generators
\begin{equation}
 \tilde \tau^{\tilde a} \partial_{\tilde a} = (A+A_i \ell^i)\partial_{\tilde u} = (A+A_i \ell^i)\partial_u,
 \label{eq:tildeti}
\end{equation}
which coincides with the translation subgroup of the initial state, i.e.
$\tilde \tau^a =\tau^a$.
The Lorentz subgroup of this final state Poincar{\' e} group has generators
\begin{equation}
\tilde \eta^{\tilde a}\partial_{\tilde a} = \frac {\tilde u}{2} \eth_{\tilde A} \tilde f^{\tilde A}\partial_{\tilde u}
      +\tilde f^{\tilde A}\partial_{\tilde A}
\end{equation}
with boosts
 \begin{equation}
\tilde B^{[i]\tilde a}\partial_{\tilde a} =-(u +\alpha)\ell^i \partial_{\tilde u} +(\eth^{\tilde A} \ell^i) \partial_{\tilde A} 
\end{equation}
and rotations
\begin{equation}
\tilde R^{\tilde a}\partial_{\tilde a} =   \epsilon^{\tilde A \tilde B}(\eth_{\tilde B} \ell^i) \partial_{\tilde A}.
\end{equation}

The Lorentz generators $\tilde B^{\tilde a}$ and $\tilde R^{\tilde a}$  can be re-expressed in terms of
the inertial coordinates $x^a = (u,x^A)$ of the initial inertial state.
The covariance of the transformation
$\tilde \xi^{\tilde a}\partial_{\tilde a} =\tilde \xi^a \partial_a $
to these initial inertial coordinates implies that $\tilde B^a$, $\tilde R^a$ and $\tilde \tau^a$
retain the Poincar{\' e} group commutation relations.
This transformation of the individual components of the final state Lorentz generators $\tilde \eta^{\tilde a}$ leads to
\begin{equation}
  \tilde R^{[i]a} \partial_a = -(R^{[i]A} \partial_A \alpha )\partial_u +R^{[i]A} \partial_A
\end{equation}
and
\begin{equation}
 \tilde B^{[i]a} \partial_a = \big( \frac{1}{2}(u+\alpha)\eth_A  B^{[i]A} -B^{[i]A}\partial_A \alpha \big )\partial_u 
     +B^{[i]A} \partial_A \, ,
\end{equation}
so that
\begin{equation}
  \tilde R^{[i]a} \partial_a = R^{[i]a} \partial_a -\epsilon^{AB}(\partial_B \ell^i)( \partial_A \alpha )\partial_u 
\label{eq:tilderr}
\end{equation}
and
\begin{equation}
  \tilde B^{[i]a} \partial_a =B^{[i]a} \partial_a  - \big( \alpha \ell^i +(\eth^A \ell^i)\partial_A \alpha \big )\partial_u .
\label{eq:tildebb}
\end{equation}
The Poincar{\' e} commutation relations, e.g. $[\tilde R^{[x]a}\partial_a,\tilde R^{[y]b}\partial_b]  =-\tilde R^{[z]c}\partial_c $,
can be checked by direct calculation.

Thus the memory effect leads to the supertranslation shift (\ref{eq:tilderr})  - (\ref{eq:tildebb}) between the components of the
preferred final and initial Lorentz groups.
For the scenario depicted in Sec.~\ref{sec:s2}, with supertranslation shift (\ref{eq:alpha}),
the rotation generators differ by
\numparts
\begin{eqnarray}
       \tilde R^{[x]a} \partial_a &=& R^{[x]a} \partial_a +\sin\phi (\partial_\theta \alpha) \partial_u\;, \\
        \tilde R^{[y]a} \partial_a &=& R^{[y]a} \partial_a -\cos\phi (\partial_\theta \alpha) \partial_u \;,\\
      \tilde R^{[z]a} \partial_a &=& R^{[z]a} \partial_a \;, 
\end{eqnarray}
\endnumparts
where
\begin{equation}
            \partial_\theta \alpha = -4m \Gamma V\sin\theta \ln  (1-V\cos\theta) \;.
 \end{equation}

\end{enumerate}

The mixing of the supertranslations with the initial and final rotation generators
leads to a mixing of
the associated supermomentum with angular momentum, whose physical
consequences have not been fully explored. Earlier accounts noted that
such supertranslation shifts could lead to a
distinctly general relativistic effect on angular
momentum conservation~\cite{BMS2,jwangular,gw}.

There are different approaches for obtaining flux-conserved
quantities which form a representations of the BMS
group. One approach consists of the linkage integrals
$L_\xi(\Sigma)$~\cite{tam,gw}, which
for each spherical cross-section $\Sigma$ of ${\mathcal I}^+$ generalize the
Komar integrals~\cite{Komar:1958wp} for exact symmetries to the case of asymptotic Killing
vectors $\xi^a$.  Associated with the linkage integrals are locally
defined fluxes ${}^L F_\xi$ whose integral determines the change
$L_\xi(\Sigma_2) - L_\xi(\Sigma_1)$ between two cross-sections.  Another
approach~\cite{dray,ashtekar-streubel,waldzoupas}
based upon the Hamiltonian phase space of gravitational
radiation also associates local fluxes ${}^H F_\xi$ and charge integrals
$Q_\xi(\Sigma)$ to each BMS generator. Here, to be specific,
we concentrate on the Hamiltonian charges
and flux because of their more physical properties.
(For a recent discussion, see~\cite{josh}.) However, the
general effect of a supertranslation shift would apply to any formulation of
BMS charges and flux.

At early and late times, denoted by $\Sigma_-$ and $\Sigma_+$, respectively,
the associated Poincar{\' e} groups lead to identical time and space translations,
i.e. $\tilde \tau^a = \tau^a$. As a result, energy-momentum conservation
takes the unambiguous form
\begin{equation}
Q_\tau(\Sigma_+) - Q_\tau(\Sigma_-) =\oint_{\Sigma_-}^{\Sigma_+}{}^H F_\tau dV \, ,
    \quad \quad  dV =\sin\theta d\theta d\phi du .
\end{equation}
However, between early and late times the associated rotation generators
differ by a supertranslation
$\tilde R^{[i]}=R^{[i]} +{\mathcal A}^{[i]}$, where ${\mathcal A}^{[i]}$ is defined
by reference to (\ref{eq:tilderr}). Consequently, the associated components
of angular momentum differ by a supermomentum,
\begin{equation}
Q_{\tilde R^{[i]}} =Q_{R^{[i]}} +Q_{{\mathcal A}^{[i]}}.
\label{eq:rdiff}
\end{equation} 
This leads to the following ambiguity in the treatment of angular momentum conservation.
The Hamiltonian flux conservation law applies to either $Q_{\tilde R^{[i]}}$ or $Q_{R^{[i]}}$,
e.g.
\begin{equation}
Q_{R^{[i]} }(\Sigma_+) - Q_{R^{[i]}} (\Sigma_-) =\oint_{\Sigma_-}^{\Sigma_+}{}^H F_{R^{[i]}}  dV .
\end{equation}
However, there {is} no flux conservation law relating the angular momentum $ Q_{\tilde R^{[i]} }(\Sigma_+)$
picked out by the final Poincar{\' e} group and the angular momentum $ Q_{ R^{[i]} }(\Sigma_i)$
picked out by the initial Poincar{\' e} group. In a typical astrophysical process leading
from initial to final non-radiative states, this complicates the interpretation of
angular momentum conservation.
\section{Discussion}

Gravitational wave memory, boosted Schwarzschild space-times and supertranslations
are three different concepts with an intriguing interconnection. The first is an astrophysical
observable corresponding to the displacement of distant masses after the passage of a gravitational wave;
the second describes a moving black hole; while the third is an asymptotical symmetry
at null infinity. In this work we have shown how the ingoing Kerr-Schild form of the
Schwarzschild metric can be used to analyze this interconnection by modeling an
idealized stationary to boosted stationary transition via a radiative stage. The model provides
a framework for approximating more realistic astrophysical systems. The chief criterion is that,
to a good approximation, the initial and final states be non-radiative and consist of the
Kerr-Schild superposition of distant Schwarzschild bodies. An example is the merger of two
initially distant stars in a quasi-Newtonian orbit to form a single moving star.
The model could even apply when the final star collapses to form a black hole.
Of course, the intermediate radiative epoch, which determines the final mass and velocity
and contributes to the null memory, must be treated by numerical methods (or some
perturbative alternative). The model presents a framework for interpreting such numerical results.

A fourth related concept is angular momentum. The difference between the
initial and final radiation strains induces a supertranslation shift between the corresponding
initial and final Poincar{\'e} groups. We have given a detailed analysis of how this supertranslation
affects the comparison of the components of the asymptotic rotation symmetries associated
with the initial and final states. As a result, the comparison of the components of angular momentum
intrinsic to the initial and final states differ by supermomenta. This complicates the interpretation
of angular momentum flux conservation laws.
{We have not attempted here to describe how radiation memory might be related to the rotational
dynamics of the sources. In the nonlinear context, this would require extensive numerical simulations.}
It is a ripe area for numerical investigation.
Computational infrastructure for computing all the BMS fluxes has been developed and
tested on simulations of the inspiral and merger 
of a precessing, spinning binary black hole~\cite{sflux}. The generalization of
our approach to the Kerr metric offers a natural framework for interpreting the
long time scale behavior of such systems of spinning bodies.

\ack
We thank the AEI in Golm  for hospitality where this project was initiated.
T.M. appreciates  support  from  C. Malone and the University of Cambridge.  
J.W. was supported by NSF grant PHY-1505965 to the University of Pittsburgh.

\end{document}